\documentstyle[mncite,draft,psfig]{mn}

\newcommand{\lta}{{\small\raisebox{-0.6ex}
{$\,\stackrel{\raisebox{-.2ex}{$\textstyle <$}}{\sim}\,$}}}

\title[Mirror eclipses in IP Peg]{Mirror eclipses in the cataclysmic 
variable IP~Peg}

\author[S.\,P.\ Littlefair et.\ al.]{S.\,P.\ Littlefair,$^{1}$
 V.\,S.\ Dhillon,$^{1}$ T.\,R.\ Marsh$^{2}$ and E.\,T.\ Harlaftis$^3$\\
$^1$Department of Physics and Astronomy, University of Sheffield, 
Sheffield S3 7RH, UK \\
$^2$Department of Physics and Astronomy, University of Southampton,
Highfield, Southampton SO17 1BJ, UK \\
$^3$Institute of Astronomy and Astrophysics, National Observatory of Athens,
P.O. Box 20048, Athens 11810, Greece\\}

\date{\center{\Large Accepted for publication in the Monthly Notices of the
Royal Astronomical Society \\ 
\vspace{.5cm} 8 June 2001}} 

\begin{document}
\maketitle

\begin{abstract} 
We present time resolved K-band infrared spectra of the dwarf nova
(DN)  IP~Peg in early quiescence. The Brackett-$\gamma$ and He-{\small
I} ($\lambda$2.0581) lines  in our data shows hitherto unseen
behaviour, which we term a {\em mirror  eclipse}, and interpret as an
eclipse of the secondary star by an optically  thin accretion disc.
Mirror eclipses are a direct probe of the structure and  physical
conditions of accretion discs. For example, on assuming the relevant
level populations to be in LTE, we constrain the temperature and
density of  the optically thin material causing the mirror eclipse in
IP~Peg to be  $10,000 \lta T \lta 20,000$K and $\rho \sim
10^{-11}$g\,cm$^{-3}$  respectively. In order to match our data, we
find that at least the outermost  20\% of the disc (in radius) must be
entirely optically thin. Implications for  time-dependent disc models
are examined.
\end{abstract} 


\section{Introduction}
\label{sec:introduction}
Cataclysmic variables (CVs) are semi-detached binary systems in which a white
dwarf primary star accretes material from a Roche-lobe filling secondary star 
via an accretion disc or a magnetically-channelled accretion flow. For a 
thorough review of CVs, see \scite{warner95a}. 

Dwarf novae (DNe) form a sub-class of the non-magnetic CVs, which are 
characterised by their quasi-periodic outbursts with amplitudes between 2 
and 6 magnitudes and intervals ranging from days to many years. These 
outbursts are generally held to be caused by an instability within the 
accretion disc -- an idea first advanced by \scite{osaki74} and later 
developed by \scite{smak84} and \scite{meyer81}. The disc instability 
hypothesis (and the models based upon it) has had considerable success in 
describing the characteristics of dwarf novae outbursts, with the 
amplitudes, periodicities, spectral evolution and basic shapes of observed 
outbursts being well-reproduced by the models \cite{cannizzo93}.

However, it has long been suspected that the discs in dwarf novae at quiescence
are quite unlike those predicted by the thermal-viscous disc instability model
(DIM). For physically viable values of the disc viscosity parameter, $\alpha$, 
the models produce discs which are optically thick. Evidence exists, however, 
which suggests that the discs in some dwarf novae are optically thin - eclipse 
mapping of quiescent DN accretion discs give disc colours which are strongly 
suggestive of optically thin gas (e.g. \pcite{wood86a}, \pcite{wood89a}). Also,
white dwarf eclipse light curves suggest an optically thin disc \cite{wood90}.
Furthermore, all versions of the DIM predict increasing fluxes from the disc 
during quiescence, whilst observations show that these fluxes stay constant or
decrease. In a recent review, \scite{lasota01} calls dwarf novae in 
quiescence the ``Achille's heel'' of the DIM.

In this paper we present time-resolved K-band spectra of the dwarf nova 
IP~Peg. IP~Peg is a deeply eclipsing system with a period of $\sim$3.8 hrs. 
It is the brightest eclipsing dwarf nova above the period gap and, as such, 
is a well-studied and important system. The Brackett-$\gamma$ and 
He-{\small I} ($\lambda$2.0581) lines in our data show hitherto unseen 
behaviour -- a reduction in equivalent width, centred on binary phase 0.5, 
which appears in the trailed spectra as a mirror-image of the classical 
rotational disturbance, or {\it Z-wave} \cite{greenstein59}.

We argue that this feature results from an eclipse of the secondary star by
an optically thin accretion disc. We model the effect in order to determine:
i) whether the optically thin region forms the main part of the disc, and ii) 
the temperature and density of the optically thin regions.

\section{Observations}
\label{sec:observations} 
On the night of 1998 August 10 and again on the nights of 2000 June 18,19
and 20 we obtained spectra of the dwarf nova IP~Peg and the M4-dwarf Gl402 with
the Cooled Grating Spectrometer 4 (CGS4) on the 3.8~m United Kingdom Infrared 
Telescope (UKIRT) on Mauna Kea, Hawaii. Both observation periods corresponded 
to early quiescence, 17-20 days after decline from outburst. CGS4 is a 1--5 
micron spectrometer containing an InSb array with 256$\times$256 pixels. 
The 40~l/mm grating with the 300~mm camera gave a resolution of 
approximately 350~km\,s$^{-1}$. To cover the wavelength range 1.81--2.51 
microns required one grating setting, centred at 2.124 microns (first order).
Optimum spectral sampling and bad pixel removal were obtained by mechanically 
shifting the array over two pixels in the dispersion direction in 
steps of 0.5 pixels.  
We employed the non-destructive readout mode of the detector to reduce
the readout noise. In order to compensate for fluctuating atmospheric 
OH$^-$ emission lines we took relatively short exposures (typically 30 
seconds) and nodded the telescope primary so that the object spectrum 
switched between two different spatial positions on the detector.
The slit width was 0.6 arcseconds (projecting to approximately 1 pixel 
on the detector) and was oriented at the parallactic angle. 
On the first run, conditions were excellent. Using the UKIRT tip-tilt 
secondary, the seeing was around $0.5''$. The humidity was low (10-20\%) 
throughout, and the sky was photometric for the duration of the run. The 
second run also had good conditions, with seeing usually around $0.5''$ and 
photometric skies. Unfortunately, a full computer disc interrupted 
observations near phase 0.5 on the first night.

A full journal of observations is presented  in table~\ref{tab:journal}. 

\section{Data Reduction} 
\label{sec:datared} 
The initial steps in the reduction of the 2D frames were performed
automatically by the CGS4 data reduction system \cite{daley94}. 
These were: the application of
the bad pixel mask, bias and dark frame subtraction, flat field division,
interlacing integrations taken at different detector positions, and co-adding
and subtracting nodded frames. Further details of the above procedures may be
found in the review by \scite{joyce92}. In order to obtain
1D data, we removed the residual sky by subtracting a polynomial fit and
then extracted the spectra using an optimal extraction technique 
\cite{horne86a}.
The next step was the removal of the ripple arising
from variations in the star brightness between integrations (i.e. at different
detector positions). These variations were due to changes in the seeing, sky
transparency and the slight motion of the stellar image relative to the slit.

There were two stages to the calibration of the spectra. The first was the
calibration of the wavelength scale using krypton arc-lamp exposures. A
fourth-order polynomial fit to the arc lines  yielded an error of less
than 0.0001 microns (rms) and the error in the fit showed no systematic
trend with wavelength. The final
step in the spectral calibration was the removal of telluric 
features and flux calibration. This was performed by dividing the spectra to 
be calibrated by the spectrum of an F-type standard star, with its prominent 
stellar features interpolated across. F-type stars were taken throughout the 
night, at different airmasses. In each case, the F-type star used was that 
which gave the best removal of telluric features, which was judged by 
inspection of the residuals. We then multiplied the result by the known flux 
of the standard at each wavelength, determined using a black body function set 
to the same effective temperature and flux as the standard. As well as 
correcting for the spectral response of the detector, this procedure also 
removed telluric absorption features from the object spectra. No correction 
for slit losses was made and the absolute flux in each spectrum is hence 
unreliable.

\section{Results}
\label{sec:results}

\subsection{Average spectra and secondary star contribution}
\label{subsec:spectra}
Figure \ref{fig:ippegav} shows the average spectra of IP~Peg from the
first run, both in our
rest frame and in the rest frame of the secondary star. The spectra of
IP~Peg in our rest frame (top) shows that the emission line of 
Brackett-$\gamma$ exhibits the double peaked line profile  
characteristic of high-inclination accretion discs. The central and lower 
spectra show that the M4-dwarf secondary star \cite{martin87} dominates the 
K-band light in IP~Peg. In order to calculate the contribution of the 
secondary star to the K-band light,  the spectra of both IP~Peg and Gl402 
were normalised and a spline fit to the continuum was subtracted. The 
spectrum of Gl402 was multiplied by a constant and subtracted from the 
spectrum of IP~Peg. This constant was then adjusted to minimise the scatter 
between the residual spectrum and a smoothed version of the residual spectrum.
We found that the secondary star in IP~Peg contributes $62\pm3$\% of the 
K-band light. 

\begin{figure}
\psfig{figure=ippegav.ps,width=8.0cm}
\caption{Average K-band spectra from the first run of the dwarf nova IP~Peg 
and the average spectrum of the M4-dwarf Gl402. The upper spectrum is the 
straight average of the time-resolved spectra of IP~Peg. The middle spectrum 
is an average of the same spectra in the rest frame of the secondary star; 
i.e. before averaging, the time-resolved spectra of IP~Peg were shifted in 
wavelength to correct for the radial velocity of the secondary star. 
Each spectrum has been normalised by dividing by the flux at 2.3 $\mu$m and 
offset by a multiple of 0.15.}
\label{fig:ippegav}
\end{figure}

\subsection{Time resolved spectra}
\label{subsec:trspectra}

\begin{figure*}
\centerline{\psfig{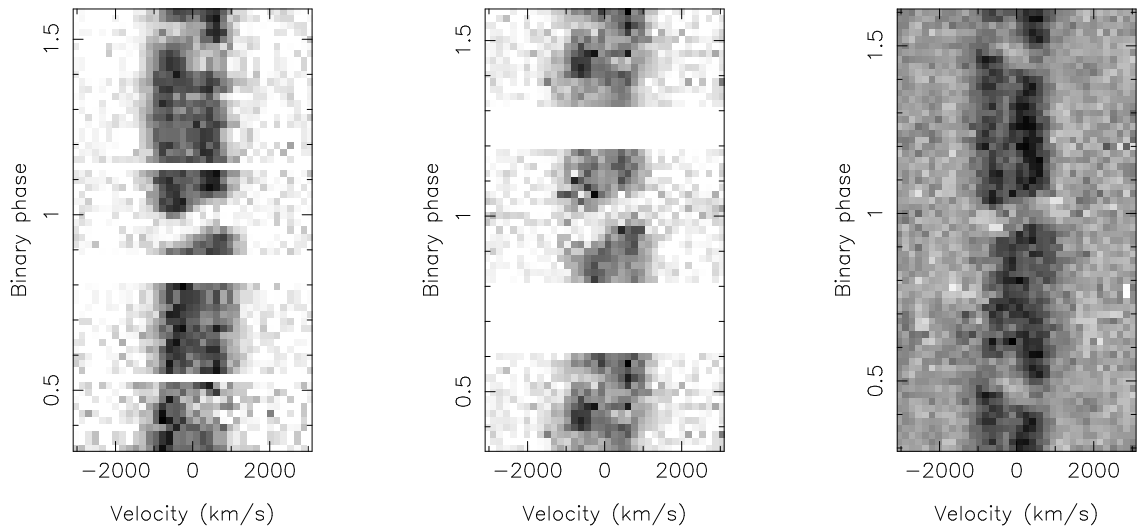}}
\caption{Time-resolved K-Band spectra of the dwarf nova IP~Peg. Dark
pixels represent strong emission. The individual
spectra have been normalised by fitting a constant to the continuum and 
dividing by this fit, and a spline fit to the continuum has been 
subtracted. Gaps show periods where observations of IP~Peg were interrupted 
to observe a calibration star. The left hand panel shows the Brackett-$\gamma$
line from the first run, taken on 1998 August 10. The middle panel shows the 
Brackett-$\gamma$ line from the second set of data, taken on the nights of 
2000 June 18,19,20. The spectra in this panel have been phase binned to 
improve signal-to-noise. The right-hand panel shows the He-{\small I} 
($\lambda$2.0581) line, where the data from all runs has been averaged into 
fifty phase bins. The middle and right-hand panels have been phase folded to 
cover the same range of phases as the left-hand panel.}
\label{fig:ippegtr}
\end{figure*}

Figure \ref{fig:ippegtr} shows the time-resolved spectra of IP~Peg in a 
trailed format.
The data were phased according to the ephemeris of \scite{beekman00}, derived
from the TiO radial velocity curves of the secondary star; the radial velocity
curves of the secondary star absorption lines in our data were found to be 
consistent with this ephemeris. 

Considering the first run (left hand panel), two ``eclipses'' are seen in 
the data, a primary eclipse centred on phase 0.0 (when the secondary star 
lies between the disc and the Earth) and a secondary ``eclipse'' centred on 
phase 0.5 and seen again after one orbital cycle at phase 1.5. The primary 
eclipse is due to the obscuration of light from the disc by the opaque 
secondary star. This eclipse is characterised by the initial eclipse of 
blue-shifted disc emission, as the secondary star first occults material from 
the disc which is moving towards the Earth. Later in the eclipse the 
receding, red-shifted part of the disc is obscured. This is the classic 
rotational disturbance pattern.

It is the secondary ``eclipse'' which is remarkable. At these phases it is 
the red-shifted, receding half of the emission line that is affected 
first. The feature possesses a telling mirror-symmetry with the classical 
rotational disturbance seen in the primary eclipse. It is this feature which 
has led us to term this a {\em mirror eclipse}. 

The mirror eclipse can be understood as follows. If the disc were not wholly 
opaque, but optically thin we would see an unusual effect. 
For a strong line, the absorption co-efficient must be higher at emission 
line wavelengths than in the continuum. Hence, more of the secondary star light
is absorbed by the disc at wavelengths corresponding to the emission lines. 
Superimposing the contributions from disc and secondary star, the absorption 
of secondary star light by the disc reduces the total amount of light at 
wavelengths corresponding to the emission lines. As the part of the disc which
initially occults the secondary star is moving away from us, we first see the 
red-shifted gas affected. During the later stages of the eclipse, the 
red-shifted emission is unaffected whilst the blue-shifted gas is now 
responsible for the absorption. This is exactly the behaviour seen during the 
mirror eclipse. Thus, a mirror eclipse is nothing more than the eclipse of the 
secondary star by an optically thin disc.

The second run (figure~\ref{fig:ippegtr} -- middle panel) also shows the 
mirror eclipse, demonstrating that the effect is persistent over long 
timescales ($\sim 2$ years). A mirror eclipse is also visible in the trailed
spectrum of He-{\small I} ($\lambda$2.0581) (figure~\ref{fig:ippegtr} - right 
hand panel.

It is not surprising that mirror eclipses have never been observed before,
as this is the first time-resolved infrared spectroscopic study of an 
eclipsing CV with a bright secondary star; the amount of secondary star light
absorbed is proportional to the intensity of the secondary star, hence 
mirror eclipses are not observed in optical spectra as the secondary stars are
much fainter (both absolutely and relative to the accretion disc) at optical 
compared to infrared wavelengths. 

\section{Modelling}
\label{sec:models}

Modelling of the mirror eclipse was undertaken with three aims in mind. 
First, we wanted to confirm that our hypothesis of an eclipse of the secondary
star by an optically thin disc could satisfactorily
reproduce the mirror eclipse as seen in the data. Second, it is possible
that the mirror eclipse effect could also be caused by a geometrically thick,
optically thin region overlying an optically thick disc. Just such a 
``sandwich-structure'' has often been proposed to explain the formation
of strong emission lines from optically thick discs \cite{mineshiga90}. 
We wished to determine if our data were compatible with such a picture. 
Third, we wanted to determine, if possible, the temperature and density of 
the gas causing the mirror eclipse.

Because our data contains no reliable flux information the data were first
normalised by fitting a constant to the continuum, then dividing the
spectra by this fit. Finally, a spline fit to the continuum around the line 
was subtracted to produce a continuum-subtracted trailed spectrum.
In order to simplify modelling, we removed the radial velocity of the white 
dwarf from the trailed spectra, and averaged the spectra which were 
unaffected by either the primary or mirror eclipse. This average spectrum was
then subtracted from the original trailed spectrum. The resulting profiles,
with the disc emission removed, represent the absorption of the secondary star
continuum by the accretion disc.

\begin{figure*}
\centerline{\psfig{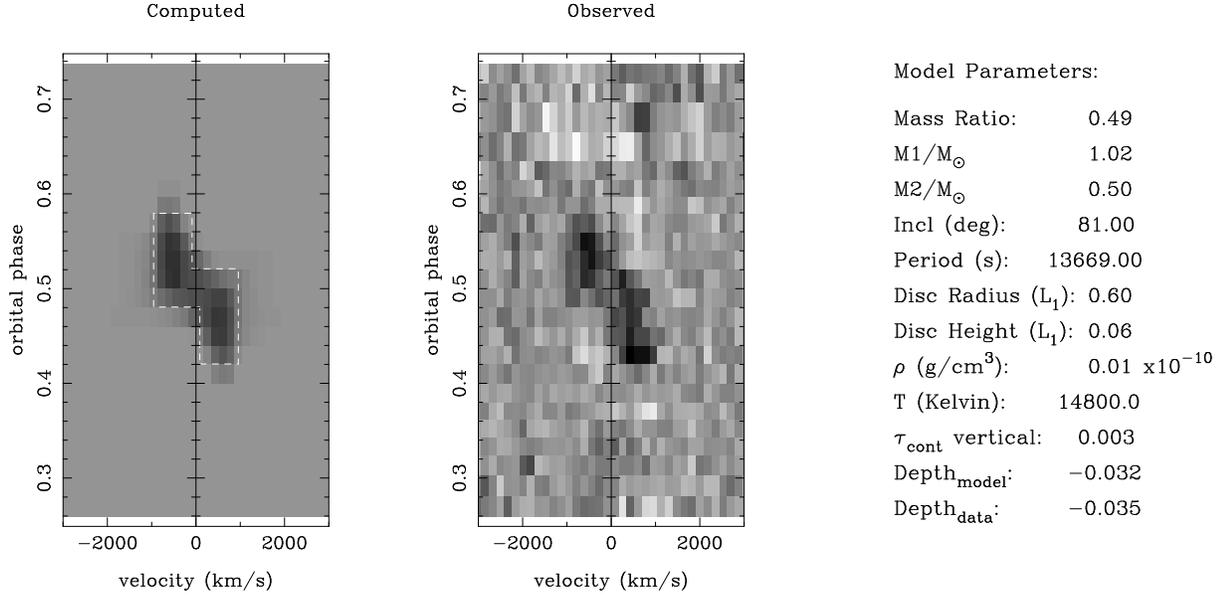}}
\caption{Results from the model described in section~\ref{sec:models}. 
Left: the computed spectra. The region used to calculate the strength of the 
mirror eclipse is shown with a dotted line. Right: the observed spectra, 
plotted on the same greyscale as the left-hand panel. The dataset shown is 
the mirror eclipse of the Brackett-$\gamma$ line. Spectra from all the runs 
have been averaged into fifty phase bins. The parameters used in the model 
are representative values for IP~Peg (table 2 in \protect\pcite{marsh90b}). 
The disc radius and height are quoted in units of distance to the inner 
Lagrangian point. In both panels, dark pixels represent strong absorption.}
\label{fig:model1}
\end{figure*}

These absorption profiles were modelled as the absorption 
of secondary star light by a non-emitting, uniform disc. The disc in our 
model is uniform in both temperature and density, the continuum opacity is 
calculated according to the formulations in \scite{gray92}. Contributions to 
the continuum opacity from H bound-free, H free-free, He-{\small I} free-free,
He-{\small I} bound-free, He$^-$ free-free and H$^-$ were considered. Line 
opacities are calculated from the line lists of \scite{hirata95}. Stark 
broadening is included for the hydrogen lines using the tables of 
\scite{lemke97}. All opacities are calculated assuming LTE and solar 
abundances. The velocity field of the disc is assumed to be Keplerian and 
vertically uniform. We take a uniform intensity distribution on the 
secondary star and trace individual rays through the disc, solving the 
radiative transfer equation to calculate the emergent line profile from a 
single element of the secondary star. The line profiles from each element are
then combined, and the resultant trailed spectrum is normalised and continuum 
subtracted in an identical fashion to the data. An estimate of the strength 
of the mirror eclipse is obtained from both data and model by calculating
the average depth within a pre-defined region of the trailed spectra. 
It is necessary to input the system parameters of 
IP~Peg into the code (masses, period and inclination). Allowing these values 
to vary within the range of values found in the literature did not 
significantly affect the results of modelling (see \scite{froning99} for
a discussion of possible system parameters for IP~Peg). 
The results are shown in figure~\ref{fig:model1}. 

We see that our simple LTE model reproduces the mirror eclipse
extremely well.  Only the disc radius, disc height, density and
temperature have been varied  to achieve agreement. The disc radius
used strongly affects both the strength and the overall shape of the
eclipse. In particular it is the disc radius  which determines the
velocity of the minima of the absorption profiles.  Disc radii of $0.5
- 0.7 L_1$ were consistent with the data -- we used a disc  radius of
$0.6L_1$, which corresponds to the disc radius found by eclipse
mapping of the infrared (H-band) continuum by \scite{froning99}.  The
disc height used was $0.06L_1$, corresponding to a  disc opening angle
of 6$^\circ$. The quiescent disc height in IP~Peg is  unconstrained by
observations; values for the opening angle in outburst vary  from
$\sim 10-20^\circ$ (\pcite{webb99}, \pcite{morales00}) and disc models
predict quiescent opening angles of $\sim 4^\circ$ \cite{hure00}.
Note that there is a strong correlation between the strength of the
mirror  eclipse and the disc height - increasing the disc height to
$0.08 L_1$ gave a  10 \% increase in mirror eclipse strength.

The model satisfactorily reproduces the strength, timing and overall 
morphology of the mirror eclipse. Hence, we are confident that our 
interpretation of the effect is correct.

\subsection{Disc structure (1) -- The line wings}
\label{subsec:struct}
There has always been a difficulty reconciling the optically thick discs 
favored by models with the presence of strong emission lines which
share the disc velocities. To overcome this, a chromosphere model has often
been suggested, in which the disc is overlaid by an optically thin
chromosphere which is the source of the emission lines, e.g. 
\scite{mineshiga90}. Further, \scite{meyer94} showed that the formation of a 
vertically extended corona is likely above a standard model disc. Both of 
these models propose a ``sandwich structure'' where optically thin material
overlays a central, optically thick region.
 
\begin{figure}
\psfig{figure=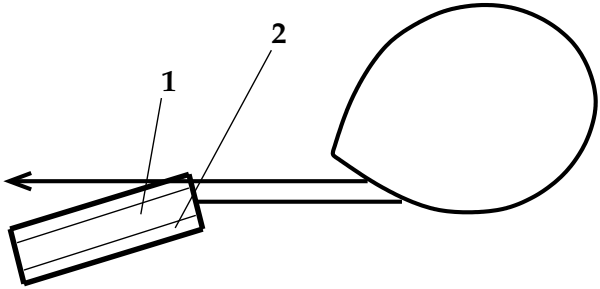,width=8.0cm}
\caption{Diagram showing how the line wings constrain geometry of the disc.}
\label{fig:diag}
\end{figure}

We might constrain such a sandwich structure by looking at the line wings 
during mirror eclipse. The line wings originate from gas with high velocities,
i.e. the inner disc. For the mirror eclipse effect to be present in the line
wings requires that light from the secondary star can follow paths which both 
pass through the inner disc regions and do not intersect any optically thick
regions of gas. An optically thick central plane would intercept rays passing 
through the inner disc. This is illustrated in figure~\ref{fig:diag}.
The disc is split into two regions. Region 1 is optically thick, and 
represents a ``standard'' model disc, region 2 is optically thin, and 
represents a chromosphere or vertically extended corona. Shown on the figure 
are two ray-paths from the secondary star to the observer. The lower ray, 
which passes through the inner disc, is absorbed by the optically
thick central plane. The upper ray passes through the optically thin 
material at larger radii. Hence, in this situation a mirror eclipse would
still be present in the line core (albeit at a lower intensity than if the
disc were entirely optically thin), but there would be no mirror eclipse in 
the line wings.
Unfortunately, there is insufficient signal-to-noise in our data to determine
if the line wings participate in the mirror eclipses exhibited by IP~Peg. 
We can, however, obtain weaker constraints on the structure of the disc by 
looking at the whole line

\subsection{Disc structure (2) -- Constraints on an optically thick midplane}
\label{subsec:struct2}
\begin{figure}
\psfig{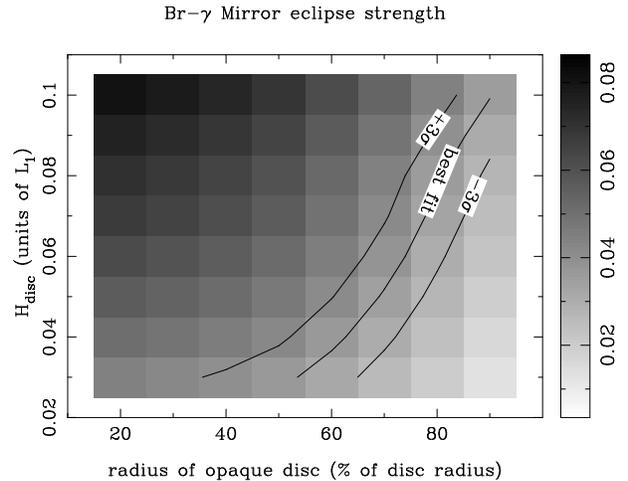}
\caption{Plot of Brackett-$\gamma$ mirror eclipse strength as a
function of  disc height and radius of an opaque obstruction located
in the disc midplane.  Mirror eclipse strength is calulated as
described in section~5, paragraph 3.  It is a dimensionless quantity.
Calculations are for a disc radius of $0.6L_1$, temperature of $6800
K$ and a density of $1.12 \times 10^{-11}$ g\,cm$^{-3}$. Contour lines
of constant mirror eclipse strength are plotted at the strength of the
Brackett-$\gamma$  eclipse in the data, and at the 3$\sigma$ levels.}
\label{fig:obstruct}
\end{figure}

Figure~\ref{fig:obstruct} shows the effect of a sandwich-structured
disc  on the strength of the Brackett-$\gamma$ mirror eclipse. Using
the code  outlined in section~\ref{sec:models}, we added an infinitely
thin, optically  thick central plane. The emergent spectrum of any ray
which intersected this  plane was set to
zero. Figure~\ref{fig:obstruct} shows the depth of the mirror eclipse
as a function of accretion disc height and the radius of the
optically thick plane. In order to determine strict upper limits on
the size  of an optically thick, central plane we used a temperature
of 6800K and a  density of $1.12\times10^{-11}$ g\,cm$^{-3}$. These
parameters give the  strongest mirror eclipse effect and so would
allow the largest possible  central obstruction. The -3$\sigma$
contour plotted shows the minimum  strength of mirror eclipse which is
consistent with our data. Models to the  right of this line cannot be
reconciled with the observations. For a disc  height of 0.06$L_1$ this
suggests that an infinitely thin, optically thick  midplane must have
a radius of less than $\sim80$\% of the disc radius.  The +3$\sigma$
contour seems to suggest that part of the disc must be opaque in order
to fit the strength of the mirror eclipse in the data. This is not the
case, however, as we have used the temperature and density which give
the strongest mirror eclipse effect. Temperatures and densities which
produce weaker mirror eclipse effects would allow a disc which is
wholly optically thin.

This result places strong limits on the extent of a sandwich structure for the
disc in IP~Peg. Even if we use those temperatures and densities which permit
the largest possible optically thick midplane and restrict the midplane itself
to be infinitely thin, we still find that the outermost 20\% of the disc 
(in radius) must be entirely optically thin. 

Note that a larger opaque central plane is permitted by invoking a
thicker  overall disc (e.\ g.\ a disc height of 0.08$L_1$ allows an
optically thick midplane with a radius of $\sim90$\% of the disc
radius, and a disc height of 0.1$L_1$ allows an optically thick
midplane with a radius of 100\% of the disc radius). The disc opening
angle in IP~Peg in outburst varies from   $\sim 10-20^\circ$
(\pcite{webb99}, \pcite{morales00}), corresponding to an outurst disc
height of $0.01-0.02 L_1$. It is therefore unlikely that  the
quiescent disc is as thick as 0.01$L_1$. Other possibilities which
allow a  thicker disc are a vertically extended chromosphere or
corona.  It is unlikely that coronal gas could be responsible  for the
mirror eclipse (see section~\ref{subsec:physical}), but a vertically
extended chromosphere remains a possibility.

\subsection{Physical conditions of the optically thin gas}
\label{subsec:physical}
\begin{figure*}
\centerline{\psfig{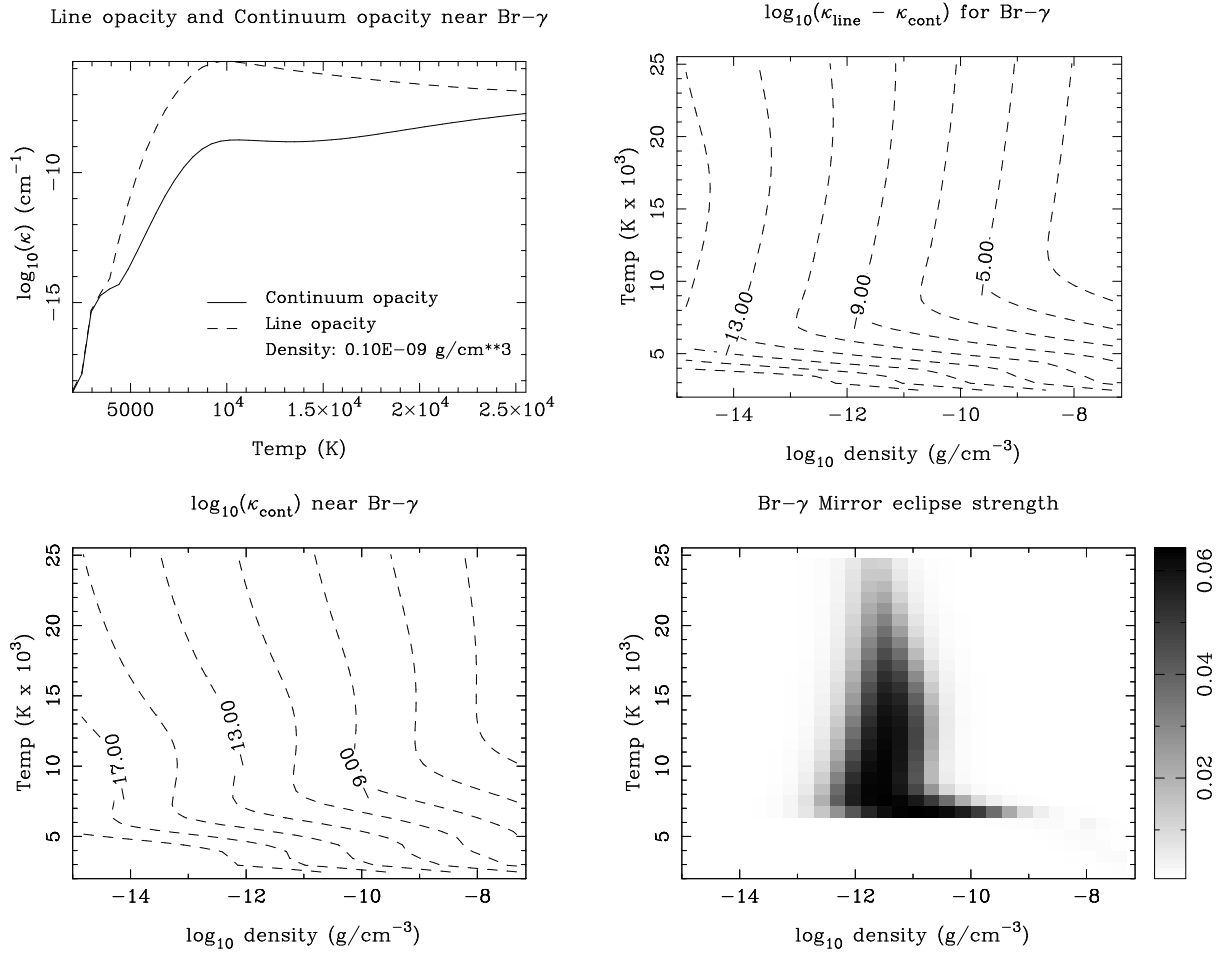}}
\caption{Top left: continuum and Brackett-$\gamma$ line opacity versus
temperature for a constant density. Bottom left: $\log_{10}$ continuum
opacity  as a function of temperature and density. Top right:
$\log_{10}$ difference  between line and continuum opacities as a
function of density and  temperature. Bottom right:  strength of
mirror eclipse in Brackett-$\gamma$ line as a function of density  and
temperature. Mirror eclipse strength is calulated as described in
section~5, paragraph 3.  It is a dimensionless quantity.  All models
assume LTE and solar abundances.}
\label{fig:brgmodels}
\end{figure*}

\begin{figure*}
\centerline{\psfig{figure=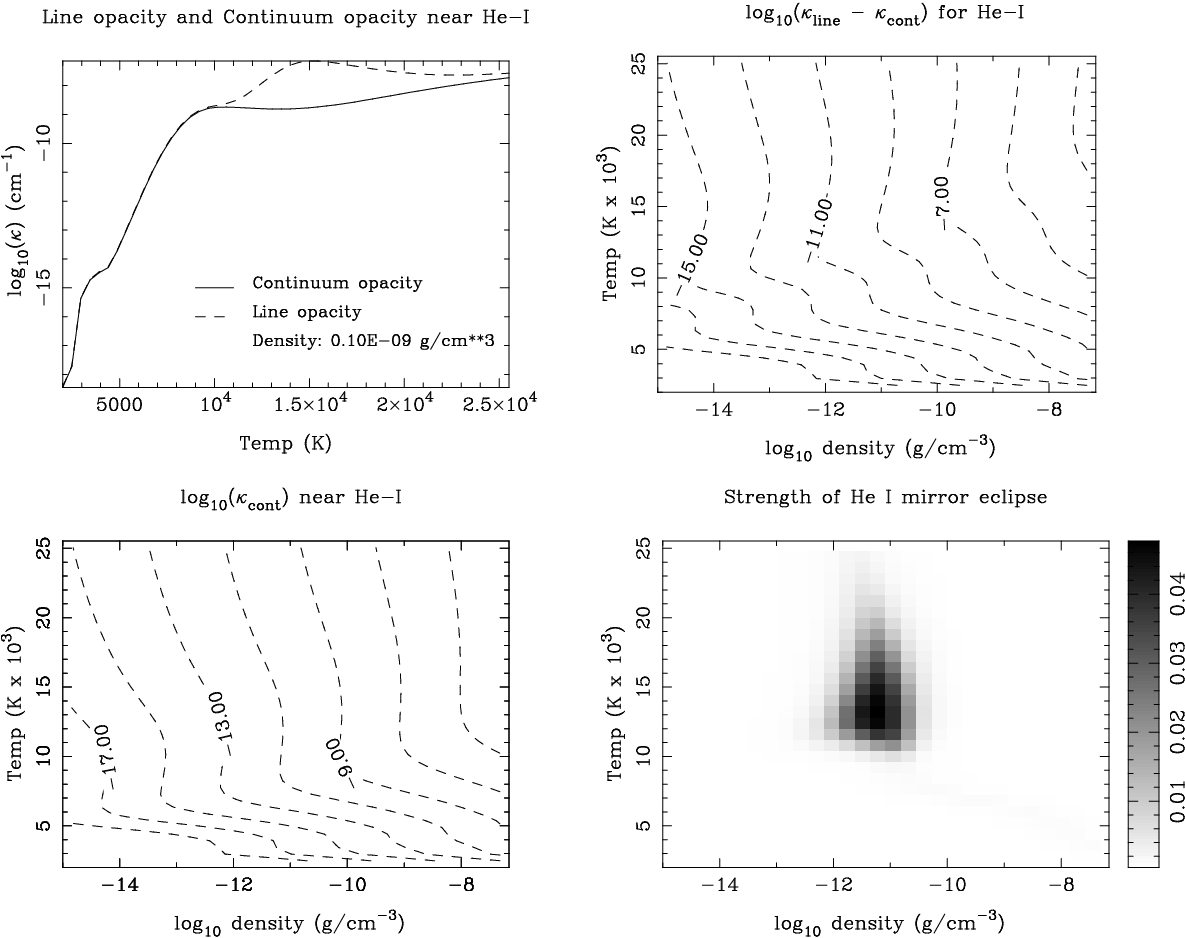,width=16.0cm}}
\caption{Top left: continuum and He {\small I} ($\lambda$2.0581) line
opacity versus temperature for a constant density. Bottom left:
$\log_{10}$  continuum opacity as a function of temperature and
density. Top right:  $\log_{10}$ difference between line and continuum
opacities as a function of  density and temperature.  Bottom right:
strength of mirror eclipse in He {\small I} ($\lambda$2.0581)  line as
a function of density and temperature. Mirror eclipse strength is
calulated as described in section~5, paragraph 3.  It is a
dimensionless quantity.  All models assume LTE and solar  abundances.}
\label{fig:heimodels}
\end{figure*}

The presence of the mirror eclipse gives constraints on the density and
temperature of the gas within the disc. This is because the mirror eclipse
requires both an optically thin continuum and a significant line opacity.
Figure~\ref{fig:brgmodels} shows how the continuum 
opacity and Brackett-$\gamma$ opacities vary with density and temperature. 
Also shown is the strength of the mirror eclipse effect over the same 
parameter space. {\em Note that no optically thick midplane was used when 
calculating this strength.} Considering the variation in the opacities with 
temperature (top-left panel), it is seen that a sharp onset of the mirror 
eclipse effect is to be expected at the temperature where the 
Brackett-$\gamma$ opacity suddenly increases. Below this temperature, although 
the disc is optically thin, there is no significant Brackett-$\gamma$ opacity 
and a mirror eclipse cannot occur. As the temperature increases further, the 
continuum opacity rises quickly and the condition that the disc is optically 
thin in the continuum is no longer satisfied. We therefore expect the mirror 
eclipse effect to be concentrated in the area close to the sudden rise of the 
line opacity. Extending our parameter space to include variations in density, 
we see that the continuum opacity increases strongly towards higher 
temperatures and densities (bottom-left panel), whilst the Brackett-$\gamma$  
opacity increases rapidly with respect to the continuum opacity along 
bow-shaped fronts (top-right panel). By comparison with the earlier example, we
would therefore expect the mirror eclipse to be concentrated along these 
fronts, weakening towards higher temperatures and densities, which is exactly 
what the model shows (bottom-right panel).

\begin{figure}
\psfig{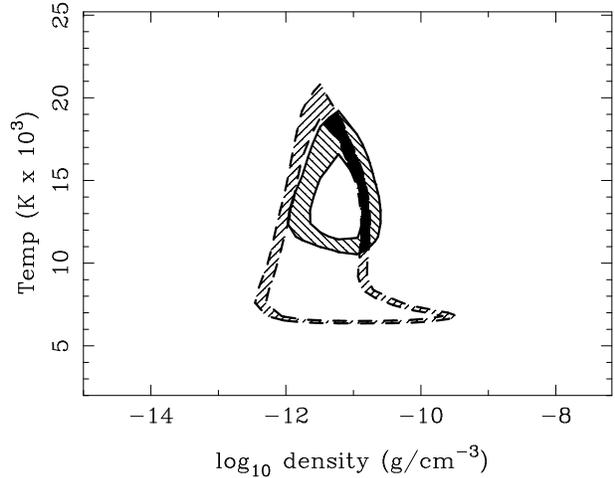}
\caption{Regions of temperature-density space which are consistent with our 
data. The solid contour bounds the region of parameter space consistent 
with the strength of the mirror eclipse in the He-{\small I} ($\lambda$2.0581) 
line. The dashed line similarly represents temperatures and densities 
consistent with the strength of the mirror eclipse in the Brackett-$\gamma$ 
line (the contour levels are determined by the strength of the mirror eclipse 
in the data, plus or minus $3\sigma$). The solid region represents 
temperatures and densities which are consistent with the observations of 
mirror eclipses in both lines.}
\label{fig:tempdens}
\end{figure}

Figure~\ref{fig:heimodels} shows the variation in continuum opacity, 
He-{\small I} ($\lambda$2.0581) line opacity and depth of mirror eclipse as a 
function of temperature and density. The mirror eclipse is confined to a 
relatively small region of parameter space (bottom-right hand panel). 
Combining this with the modelling of the Brackett-$\gamma$ mirror eclipse 
and further assuming that the same regions of gas are responsible for the 
effect in both lines (which seems reasonable given the similarities between 
the two line profiles) we can obtain rather tight constraints on the 
temperature and density of the gas responsible for the mirror eclipse. This is 
shown in figure~\ref{fig:tempdens}. 

Using constraints on the strength of the Brackett-$\gamma$ mirror eclipse in 
the data we can constrain the temperature and density of the gas causing the 
mirror eclipse to lie within the dashed contours shown in 
figure~\ref{fig:tempdens}. These contours form a band
in temperature-density space. Temperatures and densities inside the area 
bounded by this band produce a mirror eclipse which is too strong to fit the 
observed data; temperatures and densities outside the area bounded by the band 
produce too weak a mirror eclipse. Similarly, the strength of the mirror 
eclipse in He {\small I} constrains the temperatures and densities to be within
the band defined by the solid contours in figure~\ref{fig:tempdens}. The 
solid-shaded region in figure~\ref{fig:tempdens} therefore represents those 
temperatures and densities which are consistent with the strengths of the 
observed mirror eclipses in {\em both} Brackett-$\gamma$ and He-{\small I}. 

Our modelling of the conditions in the disc of IP~Peg show that the
gas responsible for the mirror eclipse has a temperature between 10,000 and 
20,000 K and a density of $\sim 10^{-11}$ g\,cm$^{-3}$ (assuming the relevant
level populations are in LTE). Brightness temperatures for the quiescent disc 
in IP~Peg have been derived by eclipse mapping in the optical and the 
infrared. Using H-band data, \scite{froning99} find a brightness temperature 
of 3000 K, whilst \scite{haswell94} use optical data to find brightness 
temperatures between 4000 and 6500 K. Since these brightness temperatures 
represent lower limits to the effective temperature of the disc, they are not
inconsistent with our findings.

These results are not easily reconciled with the DIM, where the {\em midplane}
temperature may be as low as 2000 -- 3000 K \cite{lasota01}. Certainly the
DIM predicts that the effective temperature in the disc must everywhere be
below the critical value $T_{crit} \sim 5800$ K \cite{lasota01}, nearly a 
factor of two lower than the temperatures which we find for the gas causing
the mirror eclipse. It is possible, however,  that the gas causing the mirror 
eclipse is found in an overlying chromosphere (but see 
section~\ref{subsec:struct2}). It is also possible that the high temperatures 
observed in dwarf novae accretion discs in quiescence is a consequence of the 
disc being optically thin and hence radiatively inefficient. 
\scite{williams80} found that optically thin discs possessed a ``thermostat 
effect'', their temperature never dropping below 6000~K.

The main conclusion we can draw about the midplane of the disc is that it 
must be optically thin in the K-band in its outer parts. It would be useful 
if future disc models could compute the K-band optical depth as a function of 
disc radius.

\section{Discussion}
\label{sec:discussion}
This is by no means the first evidence that the disc in IP Peg is
optically  thin at quiescence. \scite{froning99} also found evidence
for an optically  thin disc from the size of the secondary eclipse in
the H-band lightcurves, although their analysis required a complex and
uncertain decomposition of the  lightcurve. \scite{beekman00} studied
the TiO band in IP~Peg,  and found a considerable reduction in band
strength around phase 0.5.  They required a very large (approximately
the same size as the white dwarf's  Roche Lobe), optically thick disc
in order to match the strength of the band variations, in contrast to
the study by \scite{froning99}. We suggest that the difference between
these studies could  be reconciled if irradiation from the white dwarf
concentrated the TiO  absorption towards the outer hemisphere of the
secondary star (e.g.  \pcite{brett93}). Although \scite{beekman00}
found no significant ellipticity  in the radial velocity curves of
TiO, this does not necessarily imply a  uniform light distribution on
the surface of the secondary star, merely that  the light distribution
is symmetric about the line of centres. Alternatively,  if the TiO
absorption on the secondary star was concentrated in star-spot
regions, the large reduction in band strength seen by
\scite{beekman00} could  be caused by a relatively small optically
thick region, i.e. the bright spot.  Hence, we believe there is now an
overwhelming body of evidence that the  outer disc in IP~Peg is
optically thin in quiescence.

\subsection{The future of mirror eclipses}
\label{subsec:future}
Mirror eclipses have the potential to be an important tool in the study of
accretion disc physics. The presence of mirror eclipses is sensitive
to the detailed atomic physics which control line opacity.
The detection of mirror eclipses in more lines will therefore allow us to 
determine even more precisely the physical conditions of the gas within the 
disc. Mirror eclipses may lend themselves to tomography, allowing maps
of temperature and density within the disc to be made. Such extensions will
require datasets of extremely high signal-to-noise. Future studies should 
also include a study of mirror eclipses throughout an outburst cycle, allowing 
a picture of how mass builds up in the disc to be formed.

\section{Conclusions}
\label{sec:conclusions}
A {\it mirror eclipse} is an eclipse of the secondary star by an
optically thin  accretion disc. The mirror eclipse appears in the
trailed spectra as a reduction in equivalent width of the emission
lines. The mirror eclipse is velocity-dependent, with the red-shifted
part of the emission line being affected before the blue (as expected
for the sense of the rotation of the disc and binary system). This
leads to the mirror eclipse possesing a telling mirror-symmetry with
the classical rotational disturbance pattern seen in the primary
eclipse. The mirror eclipse is a consequence of a strong opacity
within the line, combined with an optically thin continuum.

As such, the observation of the mirror eclipse in IP~Peg provides clear 
evidence that at least the outer radial part of the accretion disc is optically thin. We find that an infintely thin opaque layer in the central plane of the disc cannot extend beyond about 80\% of the disc radius and still be consistent with observations.

Standard disc instability models predict optically thick discs. Modifications
to the standard disc instability models already exist which may be compatible 
with our data. Such a model has been described by \scite{gammie98}, in
which the accreted material is stored in an optically thick torus between 
outbursts. If this torus were surrounded by optically thin gas this could
be responsible for the observed mirror eclipses. 

Because the mirror eclipse is a reasonably direct probe of the physical
conditions within the disc, it provides an important
diagnostic tool, allowing determinations
of the temperature, density and structure of optically thin accretion discs.
For example, assuming the relevant level populations are in LTE, we constrain 
the temperature and density of the optically thin material causing the mirror 
eclipse in IP~Peg to be $10,000 \lta T \lta 20,000$K and 
$\rho \sim 10^{-11}$g\,cm$^{-3}$ respectively.

\section*{\sc Acknowledgements}
SPL is supported by a PPARC studentship. ETH was partially supported for
the UKIRT observing trip by the Hellenic Astronomical Society.
UKIRT is operated by the Joint Astronomy Centre on behalf of the Particle 
Physics and Astronomy Research Council. The authors acknowledge the data 
analysis facilities at Sheffield provided by the Starlink Project which is run
by CCLRC on behalf of PPARC. The authors would like to thank Rob Robinson 
and Tim Naylor for enlightening discussions and the referee and editor for
their useful comments.
\bibliographystyle{mnras}
\bibliography{refs}

\begin{table*}
\caption[]{Journal of observations. Each spectrum consists of 120 s total
exposure time, excepting those marked$^1$, which consist of 80 s total 
exposure time. Gaps between objects were used for the observation of arcs, 
flats and F-star spectra. On 18/06/00 the spectra surrounding phase 0.5 were 
lost due to a disk crash. Orbital phase is shown for IP~Peg using the 
ephemeris of \scite{beekman00}.}
\begin{center}
{\bf
\begin{tabular}{@{\extracolsep{-2.15mm}}lcccccc}
& & & & & & \\
\multicolumn{1}{l}{Object} &
\multicolumn{1}{c}{Date} &
\multicolumn{1}{c}{UT} &
\multicolumn{1}{c}{UT} &
\multicolumn{1}{c}{No. of} &
\multicolumn{1}{c}{Phase} &
\multicolumn{1}{c}{Phase} \\
 & & 
\multicolumn{1}{c}{start} &
\multicolumn{1}{c}{end} &
\multicolumn{1}{c}{Spectra} &
\multicolumn{1}{c}{start} &
\multicolumn{1}{c}{end} \\ 
& & & & & \\
IP Peg        & 10/08/98 & 09:54 & 13:12 & 78 & 0.23 & 1.08 \\ 
IP Peg        & 10/08/98 & 13:17 & 14:55 & 24 & 1.12 & 1.53 \\ 
IP Peg        & 18/06/00 & 11:50 & 13:27 & 36 & 0.26 & 0.67 \\ 
IP Peg        & 18/06/00 & 13:44 & 14:40 & 24 & 0.76 & 0.99 \\ 
IP Peg        & 18/06/00 & 14:42 & 15:10 & 12 & 1.01 & 1.12 \\ 
IP Peg        & 19/06/00 & 14:30 & 14:38 &  4 & 0.28 & 0.31 \\ 
IP Peg$^1$    & 19/06/00 & 14:38 & 15:19 & 24 & 0.32 & 0.48 \\ 
IP Peg$^1$    & 20/06/00 & 14:01 & 14:19 & 12 & 0.47 & 0.55 \\ 
& & & & & \\
\end{tabular}
}
\end{center}
\label{tab:journal}
\end{table*}

\end{document}